\documentclass{ismdproc}

\begin{document}
\title{In-medium reduction of the $\eta^\prime$ mass in $\sqrt{s_{NN}}$ = 200 GeV Au+Au collisions}


%
\author{{\slshape R\'obert V\'ertesi$^1$,
Tam\'as Cs\"org\H{o}$^{1,2}$,
J\'anos Sziklai$^1$}\\[1ex]
$^1$MTA KFKI RMKI, H-1525 Budapest 114, P.O.Box 49, Hungary\\
$^2$Dept. Physics, Harvard University, 17 Oxford Street, Cambridge, MA 02138, USA\\
}

\acronym{V\'ertesi, Cs\"org\H{o}, Sziklai -- In-medium reduction of the $\eta^\prime$ mass}

\maketitle

\begin{abstract}
  A reduction of the mass of the $\eta^\prime$(958) meson may indicate the restoration of the $U_{A}(1)$ symmetry in a hot and dense hadronic matter, corresponding to the return of the 9th, ``prodigal'' Goldstone boson. 
  We report on an analysis of a combined PHENIX and STAR data set on the intercept parameter of the two-pion 
  Bose--Einstein correlation functions, as measuremed in $\sqrt{s_{NN}} = 200\ {\rm GeV}$ Au+Au collisions at RHIC. 
  To describe this combined PHENIX and STAR dataset, an in-medium $\eta^\prime$ mass reduction of at least 200 MeV is needed, at the 99.9 \% confidence level in a broad model class of resonance multiplicities. 
  Energy, system size and centrality dependence of the observed effect is also discussed.
\end{abstract}

\section{Introduction}

Although the quark model exhibits a $U(3)$ chiral symmetry in the limit of massless up, down and strange quarks,
and in principle 9 massless Goldstone modes are expected to appear when this symmetry is broken, 
only 8 light pseudoscalar mesons are observed experimentally. This puzzling mystery is resolved by 
the Adler-Bell-Jackiw $U_{A}(1)$ anomaly: instantons tunneling between topologically different QCD vacuum 
states explicitely break the $U_{A}(1)$ part of the $U(3)$ symmetry. 
Thus the 9th Goldstone boson is expected to be massive, and is associated with 
the $\eta^\prime$ meson, which has a mass of 958 MeV, approximately twice that of the other pseudoscalar mesons.

In high energy heavy ion collisions at RHIC, a hot and dense medium is created. 
Recent measurements of the direct photon spectrum in $\sqrt{s_{NN}}$ = 200 GeV Au+Au collisions
indicate~\cite{PHENIX-directphotons}, that the initial temperature in these reactions is at least 300 MeV, 
while hadrons as we know them may not exist above the Hagedorn temperature 
of $T_H \approx 170$ MeV~\cite{Hagedorn}.  Thus the matter created
in heavy ion collisions at RHIC is hot enough to be a quark-gluon plasma~\cite{PHENIX-directphotons}. 
Detailed analysis of the properties of this matter indicate that it flows like  a perfect fluid~\cite{PHENIX-WhitePaper},
and scaling properties of the elliptic flow indicate scaling with the number of constituent quarks~\cite{PHENIX-v2scaling},
hence this matter is sometimes referred to as a strongly interacting Quark-Gluon Plasma (sQGP)~\cite{shuryak-sQGP},
or, in more direct terms, a perfect fluid of quarks~\cite{PHENIX-WhitePaper}.
          
After this perfect fluid of quarks rehadronizes, a hot and dense hadronic matter may be created, where the $U_{A}(1)$
symmetry of the strong interactions  may temporarily be restored \cite{kunihiro,kapusta,huang}.
Recent lattice QCD calculations indicate that such chirally symmetric but hadronic matter may exist
below the critical temperature for quark deconfinement~\cite{Fodor:2009ax}.
In such a medium, the mass of the $\eta^\prime(958)$ mesons may be reduced to its  quark model value of about 500 MeV, 
corresponding to the return of the ``prodigal" 9th Goldstone boson~\cite{kapusta}. 
In this presentation we summarize the the results on an indirect observation of such an in-medium $\eta^\prime$ mass modification based on a detailed analysis of PHENIX and STAR charged pion Bose-Einstein correlation (BEC) data~\cite{phnxpub,starpub}. These results have been published recently in Refs.~\cite{csvsz-PRL,Vertesi:2009ca} and detailed in Ref.~\cite{Vertesi:2009wf},

\begin{wrapfigure}{r}{0.49\textwidth}
\includegraphics[width=\linewidth]{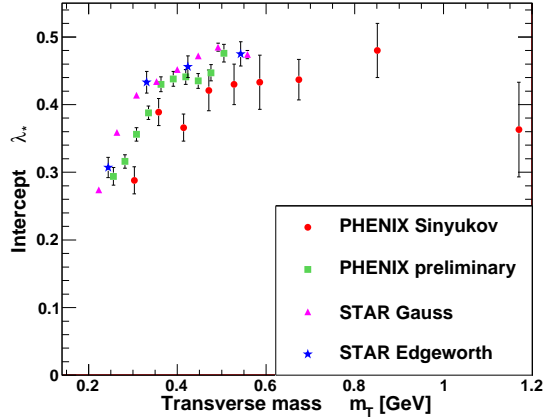}%
\caption{Datasets of $\lambda_{*}(m_{\rm T})$ from RHIC $\sqrt{s_{NN}}$ = 200 GeV like-sign pion correlation measurements from Refs.~\cite{phnxpub,phnxpre,starpub}.}
\label{f:datasets}
\end{wrapfigure}

The abundance of the $\eta^\prime$ mesons with reduced mass may be increased at low $p_{\rm T}$, 
by more than a factor of 10. One should emphasize that the $\eta^\prime$ (and $\eta$) mesons almost 
always decay after the surrounding hadronic matter has frozen out, due to their small 
annihilation and scattering cross sections, and their decay times that are much longer than the
characteristic 5-10 fm/c decoupling times of the fireball created in high energy heavy ion 
collisions. Therefore one cannot expect a direct observation of the mass shift of the $\eta^\prime$
(or $\eta$) mesons: all detection possibilities of their in-medium  mass  modification have 
to rely on their enhanced production.

An enhancement of low transverse momentum $\eta^\prime$ me\-sons contributes to an enhanced production of soft charged pions mainly through the $\eta^\prime \rightarrow \eta + \pi^+ + \pi^- \rightarrow (\pi^+ + \pi^0 + \pi^-) + \pi^+ + \pi^-$ decay chain and also through other, less prominent channels.
As the $\eta^\prime$ decays far away from the fireball, the enhanced production of pions in the corresponding halo region will reduce the strength of the Bose-Einstein correlation between soft charged pions.
The transverse mass ($m_{\rm T} = \sqrt{m^2 + p_{\rm T}^2}$) dependence of the extrapolated intercept parameter 
$\lambda_{*}$ of the charged pion Bose-Einstein correlations 
was shown to be an observable that is sensitive to such an enhanced $\eta^\prime$ multiplicity,
as pointed our first in Ref.~\cite{vance} and discussed in Ref.~\cite{Csorgo:1999sj}.  

The predicted drop of $\lambda_{*}(m_{\rm T})$ data at low transverse masses
has been observed both by PHENIX~\cite{phnxpub,phnxpre} and STAR~\cite{starpub,Abelev:2009tp} at RHIC, as it is indicated on Fig.\ \ref{f:datasets}. Note that this reduction is not present in the S+Pb data set at CERN SPS at $\sqrt{s_{NN}}$ = 19.4 GeV energy~\cite{Beker:1994qv}.

\section{Modeling and analysis method}
Our main analysis tool was a Monte-Carlo simulation of the transverse mass dependence of
the long lived resonance multiplicities including the possibility of an enhanced $\eta^\prime$
production at low transverse momentum, due to a partial in-medium $U_{A}(1)$ restoration and a related $\eta^\prime$ mass modification. 
This model and the related reduction of the effective intercept parameter of the two-pion Bose-Einstein
correlation function was proposed first in ref.~\cite{vance} and detailed recently in Refs.~\cite{csvsz-PRL,Vertesi:2009ca,Vertesi:2009wf}.

In thermal models, the production cross sections of the light mesons are exponentially suppressed by the mass. 
Hence one expects about two orders of magnitude less $\eta^\prime$ mesons from the freeze-out than pions. 
This suppression, however, may be moderated as a consequence of a possible $\eta^\prime$ mass reduction, 
and the $\eta^\prime$ mesons may show up in an enhanced number. 
The number of in-medium $\eta^\prime$ mesons is calculated with an improved Hagedorn formula
yielding the following $\eta^\prime$ enhancement factor:
\begin{equation}\label{eq:prietamtdist}
  f_{\eta^\prime}=\left(\frac{m_{\eta^\prime}^{*}}{m_{\eta^\prime}}\right)^\alpha e^{- \frac{m_{\eta^\prime}-m_{\eta^\prime}^{*}}{T_{cond}}}.
\end{equation}
This formula includes a prefactor with an expansion dynamics dependent exponent $\alpha\approx 1-d/2$ for an expansion in $d$ effective dimensions~\cite{Csorgo:1995bi}. As a default value, $\alpha = 0$ was taken~\cite{vance} and, for the systematic investigations,
this parameter was varied between $ -0.5 \le \alpha \le 0.5$.
Other model parameters and their investigated ranges  are described as follows:
$T_{cond}$ in the above formula corresponds to 
the temperature of the medium when the in-medium modified $\eta^\prime$ mesons are formed; its default value
was taken to be $T_{cond} = 177$ MeV~\cite{vance} and varied systematically
between 140 and 220 MeV. 
Resonances with different masses were simulated with a mass dependent slope parameter
$T_{eff} = T_{FO} + m \langle u_T\rangle^2 $, where the default values of $T_{FO} = 177$ MeV
and $\langle u_T\rangle = 0.48$~\cite{Adler:2003cb} were utilized and 
systematically varied in the range of 100 MeV  $\le T_{FO} \le $ 177 MeV and 0.40  $\le \langle u_T\rangle \le $ 0.60~.

Once produced, the $\eta^\prime$ is expected to be decoupled from other hadronic matter, since its annihilation and scattering cross sections are very small~\cite{kapusta}.
If the $\eta^\prime$ mass is reduced in the medium, the observed $\eta^\prime$ spectrum will consist of two components.
If the $p_{\rm T}$ of the $\eta^\prime$ is large enough, it can get on-shell and escape. 
This will produce a thermal component of the spectrum. 
Energy conservation at mid-rapidity implies ${m_{\eta^\prime}^*}^2+{p_{T,\eta^\prime}^{*}}^2={m_{\eta^\prime}}^2+{p_{T,\eta^\prime}}^2$.
(In the latter equation the quantities marked with an asterisk denote the properties of the in-medium $\eta^\prime$, while the ones without an asterisk refer to the free $\eta^\prime$.)
On the other hand, $\eta^\prime$-s with ${p_{T,\eta^\prime}^{*}} \le \sqrt{ {m_{\eta^\prime}^*}^2-{m_{\eta^\prime}}^2 }$ will not be able to leave the hot and dense region through thermal fluctuation since they cannot compensate for the missing mass~\cite{kapusta,huang}, and thus will be trapped in the hot and dense region until it disappears. As the energy density of the medium is dissolved, the effect of QCD instantons increases and the trapped $\eta^\prime$ mesons regain their free mass and appear at low $p_{\rm T}$. 

The low $p_{\rm T}$ enhancement of the $\eta^\prime$ meson also affects the spectrum of its decay products, first of all the $\eta$, the feed-down to which can be described with an $\eta^\prime$ to $\eta$ branching ratio of  $BR({\eta^\prime}\rightarrow\eta+\pi\pi) \approx 65.7 \%$.
A connection between the $\eta^\prime$ enhancement $f_{\eta^\prime}$ and the $\eta$ enhancement $f_\eta$ can be expressed as
\begin{equation}
f_\eta = 1 + \left(  f_{\eta^\prime} -1 \right)
\frac{N_{\eta^\prime}}{N_\eta}
BR({\eta^\prime}\rightarrow\eta+\pi\pi) \ ,
\end{equation}
where $N_{\eta^\prime}$ and $N_{\eta}$ denote the resonance multiplicities of the $\eta^\prime$ and $\eta$ as fixed by the input model.

In our recent works of Refs.~\cite{csvsz-PRL,Vertesi:2009ca,Vertesi:2009wf}, 
we improved on earlier simulations of Ref.~\cite{vance}, that considered the 
trapped $\eta^\prime$ mesons to leave the dissolving medium with a negligible $p_{\rm T}$ . That earlier approach 
resulted in a steep hole in the extrapolated intercept parameter 
$\lambda_{*}(m_{\rm T})$ at a characteristic transverse mass of $m_{\rm T}\le 250\ {\rm MeV}$~\cite{phnxpre,vance,Csorgo:1999sj}. 
In that simplified scenario the only free parameter was the in-medium $\eta^\prime$ mass, 
determining the depth of the observed hole. 
In our recent analysis~\cite{csvsz-PRL,Vertesi:2009ca,Vertesi:2009wf}, the $\eta^\prime$-s from the decaying condensate were given a 
random transverse momentum, following Maxwell-Boltzmann statistics with an 
inverse slope parameter $B^{-1}$, which was necessary to obtain a quality description of the 
width and the slope of the $\lambda_{*}(m_{\rm T})$ data of PHENIX and STAR in the $m_{\rm T} \approx 300$ MeV region.
Physically, $B^{-1}$ is limited by $T_{FO}$, so the trapped $\eta^\prime$ -s may gain only moderate transverse momenta.
Hence, the enhancement mostly appears at low $p_T$~\cite{kunihiro,kapusta,huang} just as in the first simulations.
However, now the slope of ``hole" of the $\lambda_{*}(m_{\rm T})$ curve is determined by $B^{-1}$,
 and, for certain values of the model parameters, the data can be reproduced quantitatively.
(The $\lambda_{*}$ values, actually used in the presented analysis, and their total errors are discussed in details in Ref.~\cite{Vertesi:2009wf}. Here $\lambda_{*}^{\rm max}$ is the $\lambda_{*}(m_{\rm T})$ value taken at $m_{\rm T}=0.7\ {\rm GeV}$, with the exception of the STAR data, where the data point at the highest $m_{\rm T}=0.55\ {\rm GeV}$ is considered. Note that the $m_{\rm T}$ dependency of the  $\lambda_{*}(m_{\rm T})$ measurements in the 0.5-0.7 GeV region is very weak.)

\begin{figure}[hb]
\begin{minipage}[t]{.49\textwidth}
\includegraphics[width=\linewidth]{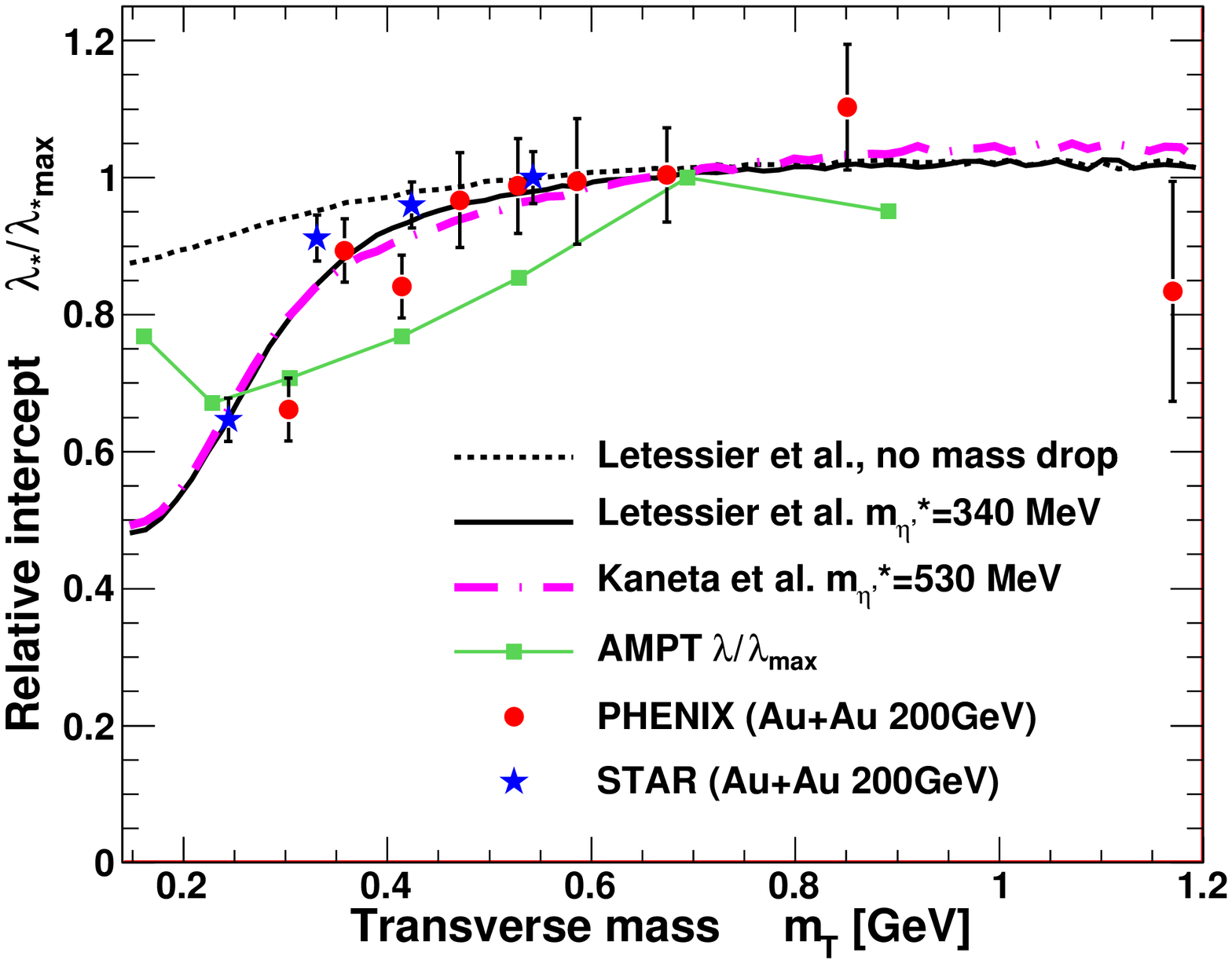}
\caption{%
The transverse mass dependence of the relative intercept parameter in the 
PHENIX and STAR dataset is reproduced with an in-medium mass modification of the
$\eta^\prime$ mesons using two different resonance models as input.
The same resonance models, but without in-medium mass modification, cannot explain these
datasets.
AMPT is known to be fairly successful in describing the HBT radii within a 
non-thermal scenario and without $\eta^\prime$ mass modification~\cite{Lin:2004en}, 
however, it is not capable of describing the current dataset in a statistically acceptable manner.
\label{f:lambdarel}
}
\end{minipage}
\hspace{\fill}
\begin{minipage}[t]{.49\textwidth}
\includegraphics[width=\linewidth]{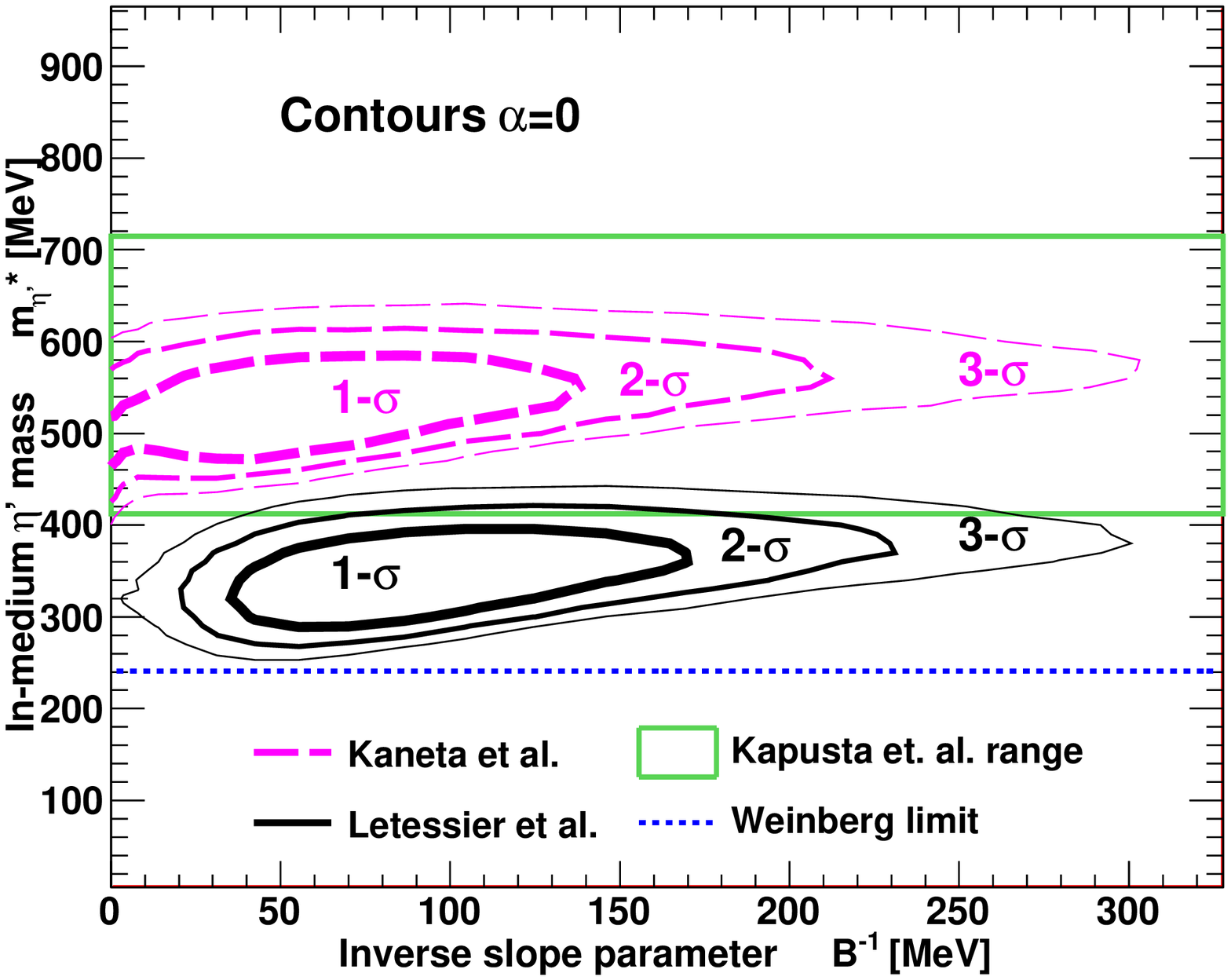}
\caption{%
  Standard deviation contours on the ($B^{-1}$, $m_{\eta^\prime}^{*}$) plain, obtained from $\lambda_{*}(m_{\rm T})/\lambda_{*}^{max}$ of Monte Carlo simulations based on particle multiplicities using two different models for hadronic resonances. The region between the horizontal solid lines indicates the theoretical range predicted by quark model considerations, while  the dotted horizontal line stands for Weinberg's lower limit.
\label{f:xcont}
}
\end{minipage}
\end{figure}

We have investigated a broad class of models of resonance production, 
including FRITIOF~\cite{fritiof} and UrQMD \cite{urqmd} models, 
that both produce resonances without assuming local thermalization.
Resonance decays, including decay chains, were simulated with JETSET 7.4~\cite{Sjostrand:1995iq}.

The FRITIOF~\cite{fritiof} Monte Carlo model, which is based on the superposition of nucleon-nucleon collisions and the Lund~string fragmentation model, cannot describe the behavior seen in $\lambda_{*}(m_{\rm T})/\lambda_{*}^{max}$ even when an arbitrary $\eta^\prime$ mass modification is considered. On the other hand, hadronic cascade based UrQMD~\cite{urqmd}, as well as the quark coalescence model ALCOR~\cite{alcor} and the thermal resonance production models of Refs.~\cite{kaneta,rafelski,stachel}, provide a successful fit in a certain range of the in-medium $\eta^\prime$ masses.
The main difference between the thermal models that we utilized was in those resonance multiplicities that are not yet measured well: ref.~\cite{kaneta} predicts a factor of 1.6 more $\eta$-s and a factor of 3 more $\eta^\prime$-s than the models of ref.~\cite{rafelski,stachel}. The relevant resonance fractions of these models are detailed in Table V of ref.~\cite{Vertesi:2009wf}.

The dotted line in Fig.~\ref{f:lambdarel} indicates a scenario without an in-medium $\eta^\prime$ mass reduction,
 while the dot-dashed and solid lines show the enhancement required to describe the dip in the low $m_{\rm T}$ region of $\lambda_{*}$ corresponding to the resonance multiplicities of Refs.~\cite{kaneta,rafelski}, respectively. 

Based on extensive Monte-Carlo simulations, $\chi^2$ of the fits to the data of Fig.~\ref{f:lambdarel}
 was computed as a function of $m_{\eta^\prime}^*$ and $B^{-1}$ for each resonance model and each fixed 
value of model parameters of $\alpha$, $T_{cond}$, $T_{FO}$ and $\langle u_{\rm T} \rangle$. 
The best values for the in-medium mass of $\eta^\prime$ mesons are in, or slightly below, the range 
$\sqrt{\frac{1}{3}(2 m_{\rm K}^2 + m_\pi^2)}\le m_{\eta^\prime}^{*} \le \sqrt{2 m_{\rm K}^2 - m_\pi^2}$
predicted in  ref.~\cite{kapusta}, while all are above the lower limit of 
$m_{\eta^\prime}^{*}\ge\sqrt{3}m_\pi$ 
given by ref.~\cite{Weinberg:1975ui}.
The $\lambda_{*}(m_{\rm T})/\lambda_{*}^{max}$ simulations for the best fits of two characteristic models are compared to the no-mass-drop scenario on Fig.~\ref{f:lambdarel}, while the 1, 2 and 3-$\sigma$ parameter boundaries are indicated in Fig.~\ref{f:xcont}. 
Models that describe both PHENIX and STAR $\lambda_{*}(m_{\rm T})/\lambda_{*}^{max}$ data in a statistically 
acceptable manner with the assumption of a sufficiently large in-medium $\eta^\prime$ mass reduction 
are all used for the estimation of systematics. The key parameters of the best fits are listed in Table~\ref{tab:modelsum}.

\section{Results }
We have used different input models and setups to map the parameter space for a twofold goal:
to determine, at least how big $\eta^\prime$ in-medium mass reduction is needed to be able to describe these datasets,
and also to determine, what are the best values of the in-medium mass modification of the $\eta^\prime$ mesons. 
Utilizing our indirect method, we have also reconstracted the 
transverse mass dependent spectrum of these $\eta^\prime$ mesons. 

\subsection{Lower limit on the in-medium $\eta^\prime$ mass reduction }
 We excluded certain regions where a statistically acceptable fit to the data is not achievable, thus we can give a lower limit on the $\eta^\prime$ mass modification. 
At the 99.9 \% confidence level, corresponding to a more than 5-$\sigma$ effect, at least 200 MeV in-medium decrease of the mass of the $\eta^\prime(958)$ meson was needed to describe both STAR 0-5 \% central and PHENIX 0-30\% central Au+Au  data on ${\lambda_{*}(m_{\rm T})/\lambda_{*}^{max}}$ in $\sqrt{s_{NN}} = 200$ GeV Au+Au collisions at RHIC, in the considered model class.

\subsection{Best value of the in-medium $\eta^\prime$ mass reduction }
We have determined the best values and errors of the fitted $m_{\eta^\prime}^{*}$ and $B^{-1}$ parameters. 
The best simultaneous description of PHENIX~\cite{phnxpub} and STAR~\cite{starpub} 
relative intercept parameter data is achieved with an $\eta^\prime$ mass that is dramatically 
reduced in the medium created in central Au+Au collisions at RHIC from its vacuum value of 958 MeV to
$340{+50\atop -60}{+280\atop -140}\pm{45}$ MeV.  The first error here is the statistical one determined by 
the 1-$\sigma$ boundaries of the fit. The second error is from the choice of the resonance model and the parameters ($\alpha$, $T_{cond}$, $T_{FO}$ and $\langle u_{\rm T} \rangle$) of the simulation. The third error is the systematics resulting from slightly different PHENIX and STAR centrality ranges, particle identification and acceptance cuts. These effects have been estimated with Monte-Carlo simulations, detailed in ref.~\cite{Vertesi:2009wf}, not to exceed 9.8\%, 7\% and 3\% respectively.
The main source of systematic errors is the choice of the resonance models. This is due to 
the unknown initial $\eta^\prime$ multiplicity, hence models like ref.~\cite{kaneta} with larger initial $\eta^\prime$ abundances require smaller in-medium $\eta^\prime$ mass modification, as compared to the models of ref.~\cite{stachel,rafelski}. 

\subsection{Transverse mass spectra of $\eta^\prime$ and $\eta$ mesons}

In addition to the characterization of the in-medium $\eta^\prime$ mass modification, 
the transverse momentum spectra of the $\eta$ and $\eta^\prime$ mesons have also been reported in
Ref.~\cite{csvsz-PRL}.

\begin{figure}[hb]
\begin{minipage}{.49\textwidth}
\includegraphics[width=\linewidth]{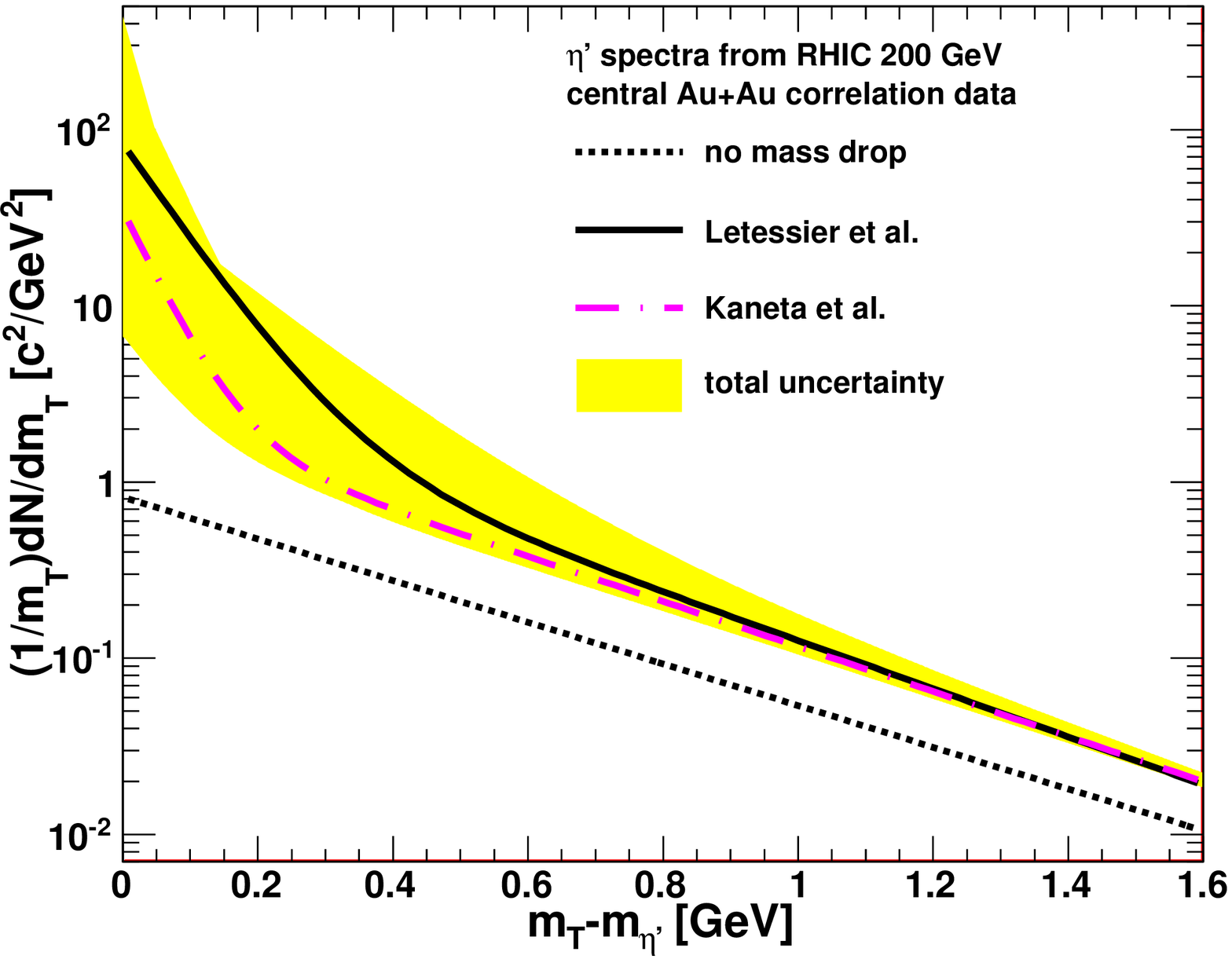}
\caption{
The transverse mass dependent spectrum  of the $\eta^\prime$ mesons,
obtained using two different resonance models as input.
The band indicates the systematic error, obtained from varying the 
resonance models as discussed in the text.
\label{f:etap-spectrum}
}
\end{minipage}
\hspace{\fill}
\begin{minipage}{.49\textwidth}
\includegraphics[width=\linewidth]{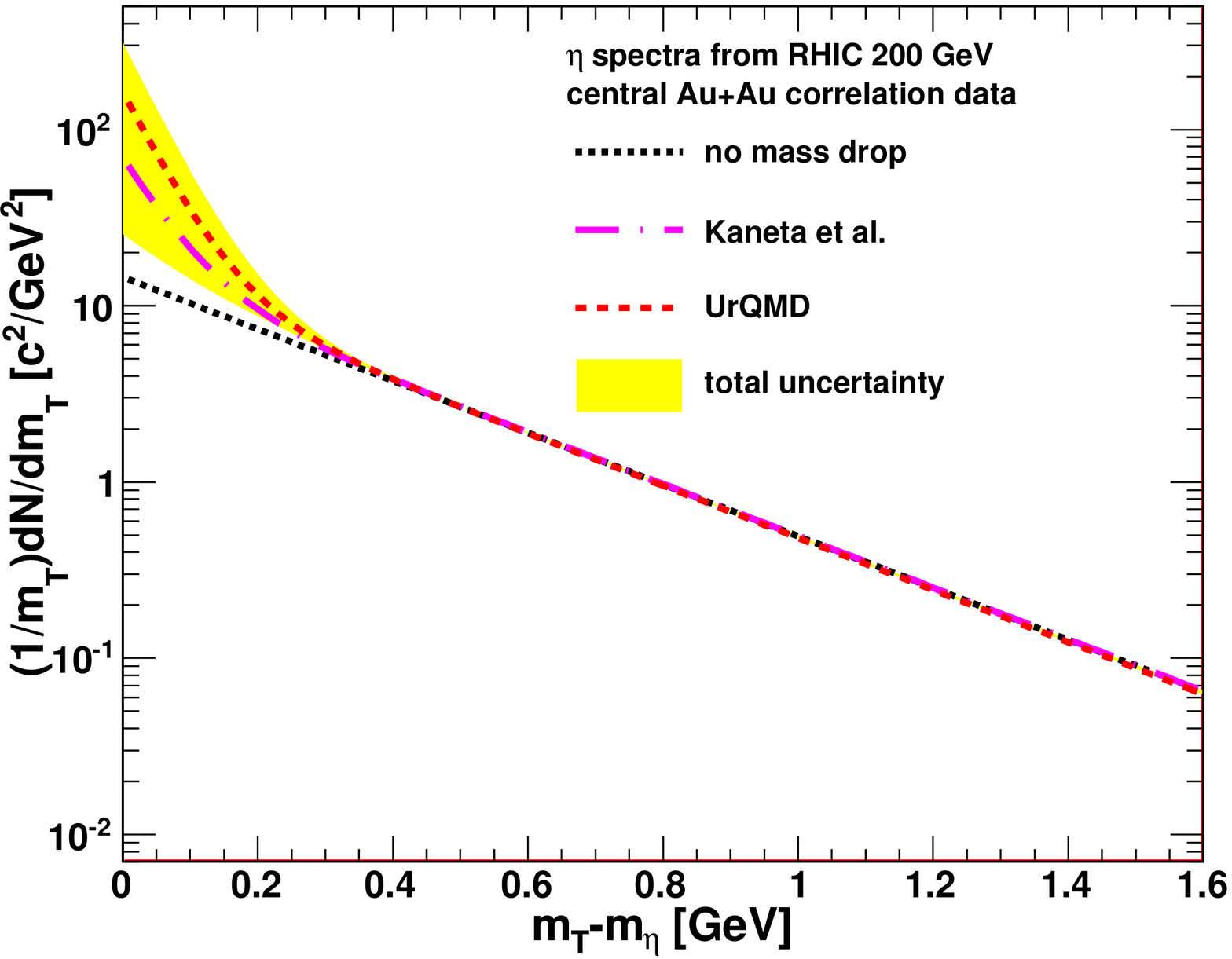}
\caption{
The transverse mass dependent spectrum of the $\eta$ mesons,
obtained using two different resonance models as input.
The band indicates the systematic error, obtained from varying the 
resonance models as discussed in the text.
\label{f:eta-spectrum}
}
\end{minipage}
\end{figure}

The reconstructed spectrum of $\eta^\prime$ and $\eta$ in $\sqrt{s_{NN}}$ = 200 GeV Au+Au collisions, shown for selected models in Figs.~\ref{f:etap-spectrum} and \ref{f:eta-spectrum} respectively, feature the characteristic low transverse momentum enhancement discussed above. 
Normalization was carried out with respect to the $\eta^\prime$ multiplicity of the model described in Ref.~\cite{kaneta}.
Although PHENIX measured before the $\eta$ spectrum in the $p_{\rm T}\ge 2\ {\rm GeV}$ region~\cite{Adler:2006bv},
as far as we know the spectrum of the $\eta^\prime$ particles has not been 
determined before in $\sqrt{s_{NN}}$ = 200 GeV Au+Au collisions at RHIC.  
Let us note that the enhancement of the $\eta$ production affects the $p_{\rm T} \le 1$ GeV region only.
The higher $p_{\rm T}$ part of the $\eta$ spectrum serves as a consistency check when 
compared to more direct measurements~\cite{Vertesi:2009wf}.

\begin{table}
\begin{center}
\begin{tabular}{|c|cccccc|c|}
\hline
Resonance & \multicolumn{6}{c|}{Best fit to combined STAR+PHENIX dataset} & 5-$\sigma$ limit on \\
model & $m_{\eta^\prime}^{*}$ (MeV)  & $B^{-1}$ (MeV) & $\chi^2 / \mbox{\it NDF}$ & CL \%
& $f_{\eta^\prime}$ & $f_\eta$ & $m_{\eta^\prime}^{*}$ (MeV) 
\\
\hline
ALCOR~\cite{alcor}
& $490{+60\atop-50}$ & 42 & 20.2/11 & 4.29 & 43.4 & 5.25 & $\le$ 700 \\
Kaneta~\cite{kaneta}
& $530{+50\atop-50}$ & 55 & 22.8/11 & 4.12 & 25.6 & 3.48 & $\le$ 730 \\
Letessier~\cite{rafelski}
& $340{+50\atop-60}$ & 86 & 18.9/11 & 6.35 & 67.6 & 4.75 & $\le$ 570 \\
Stachel~\cite{stachel}
& $340{+50\atop-60}$ & 86 & 18.9/11 & 6.38 & 67.6 & 4.97 & $\le$ 570 \\
UrQMD~\cite{urqmd}
& $400{+50\atop-40}$ & 86 & 18.9/11 & 6.14 & 45.0 & 7.49 & $\le$ 660 \\
\hline
\end{tabular}
\end{center}
\caption{%
  Best fits of $m_{\eta^\prime}^{*}$ and $B^{-1}$ for different input resonance multiplicity models, followed by $\chi^2/\mbox{\it NDF}$ and the corresponding confidence level (CL) and the integrated  $\eta^\prime$ and $\eta$ enhancement factors $f_{\eta^\prime}$ and $f_\eta$. The 5-$\sigma$ limits of maximum in-medium masses are also shown. Errors on $m_{\eta^\prime}^{*}$ values represent 1-$\sigma$ boundaries, while the 5-$\sigma$ limits include systematic errors too.
}
\label{tab:modelsum}
\end{table}

\section{Discussion}

The mass drop effect has been inspected from several aspects including the dependence on the properties of the colliding systems. Note however, that the validity of our analysis relies on the correctness of published STAR and PHENIX data. These data, however, were not measured with the definite purpose to serve as a base for the search for the in-medium $\eta^\prime$ mass modification, where particular attention has to be payed to the momentum dependence of the particle identification purity and efficiency, especially at low $p_{\rm T}$ regime.
More detailed dedicated $\lambda_{*}(m_{\rm T})$ measurements together with additional analysis of the dilepton and photon decay channels of $\eta^\prime$ could help consolidate the findings reported here.

\subsection{System size, centrality and energy dependence}

\begin{wrapfigure}{r}{0.49\textwidth}
\includegraphics[width=\linewidth]{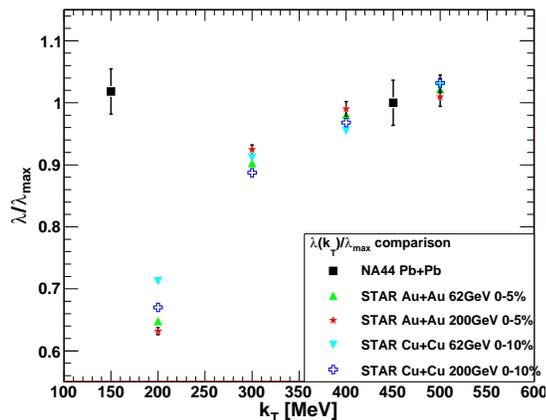}
\caption{%
The energy and system size dependence of the relative intercept parameter in the 
NA44 S+Pb and STAR Cu+Cu and Au+Au datasets.
\label{f:lambdarel-excite}
}
\end{wrapfigure}

Detailed analysis of the STAR and PHENIX $\lambda_{*}(m_{\rm T})/\lambda_{*}^{max}$ dataset recorded at 7.7, 9.2, 11.5, 39 and 62.4 GeV during 2010 has just been started~\cite{Abelev:2009bw}, marking the beginning of the RHIC energy scan program.

Detailed data available from the NA44 collaboration at $\sqrt{s_{NN}}=19.4\ {\rm GeV}$~\cite{Beker:1994qv} as well as from the STAR collaboration at different centralities within the 0\%--80\% range at $\sqrt{s_{NN}}=62.4$ and 200 GeV Cu+Cu and Au+Au collisions~\cite{Abelev:2009tp} is shown on 
Fig.\ \ref{f:lambdarel-excite}.
The NA44 data at $\sqrt{s_{NN}}=19.4\ {\rm GeV}$ does not feature an $\eta^\prime$ mass drop effect. However a positive sign of the $\eta^\prime$ mass modification is apparent in each case of the STAR datasets, indicating that the mass modification effect is nearly at maximum in $\sqrt{s_{NN}}=200\ {\rm GeV}$ Au+Au collisions and reduces with decreasing centrality, colliding energy and system size. We have estimated the magnitude of the system size and energy dependence between 62.4 GeV Cu+Cu and 200 GeV Au+Au collisions to be not larger than 15\%, which is substantially less than the dominant systematic error coming from the choice of the resonance model. 

\subsection{Channels other than BEC}

A promising channel of observation is the dileptonic decay $\eta^\prime\rightarrow\ell^+\ell^-\gamma$, because a low-$p_{\rm T}$ $\eta^\prime$ enhancement would give extra lepton pairs to the low invariant mass region. The paper of Kapusta, Kharzeev and McLerran on the return of the prodigal Goldstone boson~\cite{kapusta} was in fact motivated by the dilepton enhancement seen in CERN SPS 200 AGeV/c energies in S+Pb reactions.
Recent interpretations of CERES~\cite{ceresll} and NA60 data~\cite{na60ll} indicate that the approach
to a chiral symmetry restored state could proceed through resonance broadening and eventually 
subsequent melting, rather than by dropping masses or mass dependency or mass degeneracy between 
chiral partners~\cite{Tserruya:2006ht}.
Recent PHENIX findings also show a definite excess in the $m_{{\rm e}^+{\rm e}^-}\lesssim 1\ {\rm GeV}$
dielectron invariant mass region in $\sqrt{s_{NN}}$ = 200 GeV Au+Au collisions~\cite{Adare:2009qk}. 
Unlike at lower beam energies, in this case the contribution from a hot hadronic phase without mass 
shifts seems to be insufficient to account for the enhancement seen in the data~\cite{Drees:2009xy}.
Estimations using the enhancement factors in Table~\ref{tab:modelsum} indicate that the observed in-medium $\eta^\prime$ mass drop is indeed a promising candidate to explain this dilepton excess.

PHENIX recently reported a two-component transverse momentum spectrum in dilepton channel {\it direct photon measurements}~\cite{Adare:2009qk}, which provides an additional testing possibility to constrain the two component structure of the $\eta^\prime$ spectra reported here.

\section{Summary}
Our report presents a statistically significant, 
indirect observation of an in-medium mass modification 
of the $\eta^\prime$ me\-sons in $\sqrt{s_{NN}} = 200$ GeV Au+Au collisions at RHIC.
These results were recently published in Refs.~\cite{csvsz-PRL,Vertesi:2009ca,Vertesi:2009wf}.
A similar search for in-medium $\eta^\prime$ mass modification provided negative result in S+Pb reactions at
CERN SPS energies~\cite{vance}.  
More detailed studies of the excitation function, 
the centrality and system size dependence of the relative intercept $\lambda_{*}(m_{\rm T})/\lambda_{*}^{max}$ could provide important additional 
information about the onset and saturation of the partial $U_{A}(1)$ symmetry restoration in hot and dense hadronic matter. 
Studies of the low-mass dilepton spectrum and measurements of other decay channels 
of the $\eta^\prime$ meson may shed more light on the reported magnitude of the low $p_{\rm T}$ $\eta^\prime$ enhancement 
and the related $U_{A}(1)$ symmetry restoration in high energy heavy ion collisions.

\section*{Acknowledgments}
We thank the Organizers of the ISMD 2010 and the HCBM 2010 conferences for creating inspiring scientific atmospheres 
and providing excellent settings for scientific discussions. We also would like to thank to professors R.~J.~Glauber and Gy.~Wolf for inspiring and clarifying discussions. 
T.~Cs. is grateful to R.~J.~Glauber for his kind hospitality at the Harvard University and also to the organizers of ISMD 2010 for the kind invitation.
R.~V. is thankful to the organizers of the HCBM 2010 workshop for the invitation.
Our research was supported by Hungarian OTKA grant NK 73143. T. ~Cs. has also been supported by
a Senior Leader and Scholar Fellowship by the Hungarian American Enterprise Scholarship Fund (HAESF).
 

\newpage

\begin{footnotesize}

\end{footnotesize}



\begin{thebibliography}{99}

\bibitem{PHENIX-directphotons}
  A.~Adare {\it et al.}  [PHENIX Collaboration],
  Phys.\ Rev.\ Lett.\  {\bf 104}, 132301 (2010)
  [arXiv:0804.4168 [nucl-ex]].

\bibitem{Hagedorn}
  R.~Hagedorn,
  Nuovo Cim.\ Suppl.\  {\bf 3}, 147 (1965).

\bibitem{PHENIX-WhitePaper}
  K.~Adcox {\it et al.}  [PHENIX Collaboration],
  Nucl.\ Phys.\  A {\bf 757}, 184 (2005)
  [arXiv:nucl-ex/0410003].

\bibitem{PHENIX-v2scaling}
  A.~Adare {\it et al.}  [PHENIX Collaboration],
  Phys.\ Rev.\ Lett.\  {\bf 98}, 162301 (2007)
  [arXiv:nucl-ex/0608033].


\bibitem{shuryak-sQGP}
  E.~Shuryak,
  Nucl.\ Phys.\  A {\bf 774}, 387 (2006)
  [arXiv:hep-ph/0510123].


\bibitem{kunihiro} 
  T.~Kunihiro,
  Phys.\ Lett.\  B {\bf 219}, 363 (1989); 
  ibid. {\bf 245} 687(E) (1990).


\bibitem{kapusta} 
  J.~I.~Kapusta, D.~Kharzeev and L.~D.~McLerran,
  Phys.\ Rev.\  D {\bf 53}, 5028 (1996).

\bibitem{huang} 
  Z.~Huang and X.~N.~Wang,
  Phys.\ Rev.\  D {\bf 53}, 5034 (1996).

\bibitem{Fodor:2009ax}
  Z.~Fodor and S.~D.~Katz,
  arXiv:0908.3341 [hep-ph].

\bibitem{phnxpub} 
  S.~S.~Adler {\it et al.}, 
  Phys.\ Rev.\ Lett.\  {\bf 93}, 152302 (2004).

\bibitem{phnxpre} M. Csan\'ad for the PHENIX Collaboration, Nucl. Phys. A {\bf 774} 611-614 (2006).

\bibitem{starpub} 
  J.~Adams {\it et al.}, 
  Phys.\ Rev.\  C {\bf 71}, 044906 (2005).

\bibitem{csvsz-PRL}
  T.~Cs\"org\H{o}, R.~V\'ertesi and J.~Sziklai,
Phys.\ Rev.\ Lett.\  {\bf 105}, 182301 (2010)
  [arXiv:0912.5526 [nucl-ex]].

\bibitem{Vertesi:2009ca}
  R.~V\'ertesi, T.~Cs\"org\H{o} and J.~Sziklai,
  Nucl.\ Phys.\  A {\bf 830}, 631C (2009).

\bibitem{Vertesi:2009wf}
  R.~V\'ertesi, T.~Cs\"org\H{o} and J.~Sziklai,
  arXiv:0912.0258 [nucl-ex].

\bibitem{vance} 
  S.~E.~Vance, T.~Cs\"org\H{o} and D.~Kharzeev, 
  Phys.\ Rev.\ Lett.\  {\bf 81}, 2205 (1998).

\bibitem{Csorgo:1999sj}
  T.~Cs\"org\H{o},
  Heavy Ion Phys.\  {\bf 15}, 1 (2002)
  [arXiv:hep-ph/0001233].

\bibitem{Abelev:2009tp}
  B.~I.~Abelev {\it et al.},
  Phys.\ Rev.\  C {\bf 80}, 024905 (2009).

\bibitem{Beker:1994qv}
  H.~Beker {\it et al.},
  Phys.\ Rev.\ Lett.\  {\bf 74}, 3340 (1995).

\bibitem{Csorgo:1995bi}
  T.~Cs\"org\H{o} and B.~L\"orstad,
  Phys.\ Rev.\  C {\bf 54}, 1390 (1996).

\bibitem{Adler:2003cb} S.~S.~Adler {\it et al.}, Phys.\ Rev.\ C {\bf 69}, 034909 (2004).

\bibitem{fritiof} B. Anderson {\it et al.}, Nucl. Phys. B {\bf 281}, 289 (1987).

\bibitem{urqmd} M.~Bleicher {\it et al.}, J.\ Phys.\ G {\bf 25}, 1859 (1999).

\bibitem{Sjostrand:1995iq} 
  T.~Sj\"ostrand, 
  Comp.\ Phys.\ Commun. {\bf 82}, 74 (1994).

\bibitem{alcor}  T.~S.~Bir\'o, P.~L\'evai and J.~Zim\'anyi, Phys.\ Lett.\  B {\bf 347}, 6 (1995).

\bibitem{kaneta} M. Kaneta and N. Xu, arXiv:nucl-th/0405068.

\bibitem{rafelski} J.~Letessier, J.~Rafelski, Eur.\ Phys.\ J.\  A {\bf 35}, 221 (2008).

\bibitem{stachel} S.~A.~Bass {\it et al.}, Nucl.\ Phys.\  A {\bf 661}, 205 (1999).

\bibitem{Lin:2004en}
  Z.~W.~Lin, C.~M.~Ko, B.~A.~Li, B.~Zhang and S.~Pal,
  Phys.\ Rev.\  C {\bf 72}, 064901 (2005).

\bibitem{Weinberg:1975ui}
  S.~Weinberg,
  Phys.\ Rev.\  D {\bf 11}, 3583 (1975).

\bibitem{Adler:2006bv}
  S.~S.~Adler {\it et al.},
  Phys.\ Rev.\  C {\bf 75}, 024909 (2007).

\bibitem{Abelev:2009bw}
  B.~I.~Abelev {\it et al.},
  Phys.\ Rev.\  C {\bf 81}, 024911 (2010).

\bibitem{ceresll} A.~Marin {\it et al.}  [CERES Collaboration], PoS CPOD07, 034 (2007).
\bibitem{na60ll} E.~Scomparin  [NA60 Collaboration], PoS CPOD07, 033 (2007).
\bibitem{Tserruya:2006ht} I.~Tserruya,  Nucl.\ Phys.\  A {\bf 774}, 415 (2006).

\bibitem{Adare:2009qk}
  A.~Adare {\it et al.},
  Phys.\ Rev.\  C {\bf 81}, 034911 (2010).

\bibitem{Drees:2009xy} A.~Drees, arXiv:0909.4976 [nucl-ex].

\end{thebibliography}
\end{document}